\documentstyle[emulate_apj,apjfonts,epsf]{article}
\gdef\h50min{$h_{50}^{-1}$}
\gdef\kms{km\,s$^{-1}$}
\gdef\j0848{RX\,J0848+4453}
\lefthead{van Dokkum et al.}
\righthead{Galaxy Population of \j0848}
\slugcomment{Accepted for publication in The Astrophysical Journal Letters}
\begin{document}

\title{The Galaxy Population of Cluster \j0848 at $z=1.27$
\altaffilmark{1}}

\author{P. G. van Dokkum\altaffilmark{2,3},
S. A. Stanford\altaffilmark{4,5}, B. P. Holden\altaffilmark{4,5},
P. R.\ Eisenhardt\altaffilmark{6}, M.\ Dickinson\altaffilmark{7}, \&
R.\ Elston\altaffilmark{8}}

\altaffiltext{1}{Based on observations with the NASA/ESA {\em Hubble Space
Telescope}, obtained at the Space Telescope Science Institute, which
is operated by AURA, Inc., under NASA contract NAS 5--26555.}
\altaffiltext{2}{Hubble Fellow}
\altaffiltext{3}{California Institute of Technology, MS105-24, Pasadena,
CA 91125}

\altaffiltext{4}{Physics Department, University of California-Davis, Davis, CA 
95616, USA}

\altaffiltext{5}{Institute of Geophysics and Planetary Physics, Lawrence Livermore National Laboratory}

\altaffiltext{6}{Jet Propulsion Laboratory, California Institute of
Technology, MS 169-327, 4800 Oak Grove, Pasadena, CA 91109}

\altaffiltext{7}{Space Telescope Science Institute, 3700 San Martin Drive, Baltimore, MD 21218}

\altaffiltext{8}{Department of Astronomy, University of Florida, P.O. Box 112055, Gainesville, FL 32611}

\begin{abstract}

We present a study of the galaxy population in the IR-selected cluster
\j0848 at $z=1.27$, using deep {\em Hubble Space Telescope} (HST)
NICMOS $H_{F160W}$ and WFPC2 $I_{F814W}$ images of the cluster core.
We morphologically classify all galaxies to $K_s=20.6$ that are
covered by the HST imaging, and determine photometric redshifts using
deep ground based $BRIzJK_s$ photometry.  Of 22
likely cluster members with
morphological classifications, eleven (50\,\%)
are classified as early-type
galaxies, nine (41\,\%) as spiral galaxies, and two (9\,\%)
as ``merger/peculiar''.
At HST resolution the second brightest cluster galaxy
is resolved into a spectacular merger between three red galaxies of
similar luminosity, separated from each other
by $\approx 6$\,\h50min\,kpc, with an integrated magnitude
$K=17.6$ ($\sim 3 L_*$ at $z=1.27$). The two
most luminous early-type galaxies
also show evidence for recent or ongoing interactions.
Mergers and interactions between galaxies
are possible because \j0848{} is not yet relaxed.
The fraction of early-type galaxies in our sample
is similar to that in clusters at $0.5<z<1$, and consistent with
a gradual decrease of the number of early-type galaxies in clusters
from $z=0$ to $z \approx 1.3$.
We find evidence that the color-magnitude relation 
of the early-type galaxies
is less steep than in the nearby Coma cluster.
This may indicate that the brightest early-type galaxies have
young stellar populations at $z=1.27$, but is also consistent
with predictions of single age ``monolithic'' models with a galactic wind.
The scatter in the color-magnitude relation is $\approx 0.04$
in rest frame $U-V$, similar to that in clusters at $0<z<1$.
Taken together, these results show that 
luminous early-type galaxies
exist in clusters at $z \approx 1.3$, but that their number density
may be smaller than in the local Universe. Additional
observations are needed to determine whether the brightest early-type
galaxies harbor young stellar populations.

\end{abstract}


\section{Introduction}

Studies of nearby and distant clusters have provided strong
constraints on the evolution and formation of their galaxy population.
Approximately 80\,\% of galaxies in the central regions of nearby
clusters are elliptical and S0 galaxies (Dressler 1980).  Studies of clusters
at $0<z<1$ have shown that the luminosities and colors of these
early-type galaxies evolve slowly with time, and have very small
scatter (e.g., Schade et al.\ 1996, Ellis et al.\ 1997, Stanford,
Eisenhardt, \& Dickinson 1998, van Dokkum et al.\ 1998, De Propris et
al.\ 1999), indicating that the stars in early-type
galaxies were formed at $z > 2$.

On the other hand, there is growing evidence for morphological
evolution among cluster galaxies.  Dressler et al.\ (1997) reported a
high fraction of spiral galaxies in clusters at $0.3<z<0.5$.  Other
studies have confirmed this trend, and extended it to $z=0.8$ (Couch
et al.\ 1998; van Dokkum et al.\ 2000). These spiral galaxies
may be the progenitors of S0 galaxies
in nearby clusters (e.g., Butcher \& Oemler 1984,
Dressler et al.\ 1997). Furthermore,
rich clusters at $0.3<z<0.9$ show enhanced
rates of interactions and mergers (e.g., Couch et al.\ 1998,
van Dokkum et al.\ 1999), which may lead to the formation of
young elliptical galaxies at late times.
As a result of these processes,
early-type galaxies in distant
clusters may form a subset of all progenitors of early-type galaxies
in nearby clusters (e.g., van Dokkum \& Franx 2001).

The strongest constraints on the formation and evolution of cluster
galaxies come from the highest redshift data. 
Here, we present a
study using NICMOS and WFPC2 on HST of the galaxy population in
\j0848 at $z=1.27$. The cluster was
discovered by Stanford et al.\ (1997), as part of a deep $BRIzJK_s$ field
survey (Eisenhardt et al.\ 2001).  Its redshift corresponds to a
time when the Universe was only $35$\,\% of its present age,
and the properties of its galaxy population provide
direct information on the early stages of the formation of cluster
galaxies. We assume $\Omega_m=0.3$, $\Omega_{\Lambda}=0$,
and $H_0=50$\,\kms\,Mpc$^{-1}$.

\section{Data}

On 1 March 1999 the cluster was observed by WFPC2 on HST.  Ten
exposures were obtained in the $I_{F814W}$ band for a total of 27.8 ks.  The
exposures were reduced using standard procedures and combined with
``drizzle'' (Fruchter \& Hook 1997).

A mosaic of three NIC3 pointings in $H_{F160W}$ was obtained on 5 and
6 June 1998 during a campaign when the HST secondary was moved to
ensure optimal focus for Camera 3. 
Each pointing consisted of 8 dithered
exposures each lasting $\sim$1400~s.  The data were processed using a
combination of STScI pipeline routines and custom software, and were
combined into a single mosaic using drizzle by
registering to the WFPC2 reduced image. The PSF of the drizzled
NICMOS data has FWHM\,$\sim 0.22$\arcsec. The overlap of the two
datasets is good but not complete; results presented below will make
use of objects for which we have either WFPC2 or NIC3 data, or both.

Ground-based imaging data obtained as previously reported in
Stanford et al.\ (1997) were used in calculating photometric redshifts 
as described below.  To summarize: deep imaging was obtained at
the KPNO 4~m using PFCCD in the $BRIz$ bands and IRIM in the $JK_s$
bands over a $\sim 5\arcmin \times 5\arcmin$ area in Lynx (Eisenhardt et
al.\ 2001).  Within this area, deep $JHK_s$ band data
which cover $\sim 2.5\arcmin \times 2.5\arcmin$ were also obtained
reaching e.g.\ a 4$\sigma$ limit of $K_s \sim 21.7$.  

\section{Photometric Redshifts}

Ideally, the galaxy population of clusters is studied using magnitude
limited samples of spectroscopically confirmed members (e.g., Lubin et
al.\ 1998). However, even using the Keck Telescopes only
$10$ member redshifts have
been obtained within the HST field, due to the faintness
of the galaxies (Stanford et al.\ 1997 and
in preparation).  Therefore we chose to determine
photometric redshifts for all galaxies in the fields covered by NIC3
and/or WFPC2 to a limiting magnitude of $K_s=20.6$.  This approach has
the advantage of a uniform selection in a red filter, at the expense
of having a small amount of contamination by field galaxies.

We used the publically available ``hyperz'' code by Bolzonella,
Miralles, \& Pello (2000)
to determine photometric redshifts, using the $BRIzJK_s$
ground based photometry.  Empirical templates from Coleman, Wu,
\& Weedman (1980) were
used in the SED fitting. The assigned redshifts are weighted
averages of fits to four spectral types ranging from E/S0 to a star
burst spectrum.  Figure \ref{zhis.plot} shows the resulting
photometric redshift distribution of galaxies in the HST
field. 

\vbox{
\begin{center}
\leavevmode
\hbox{%
\epsfxsize=7.2cm
\epsffile{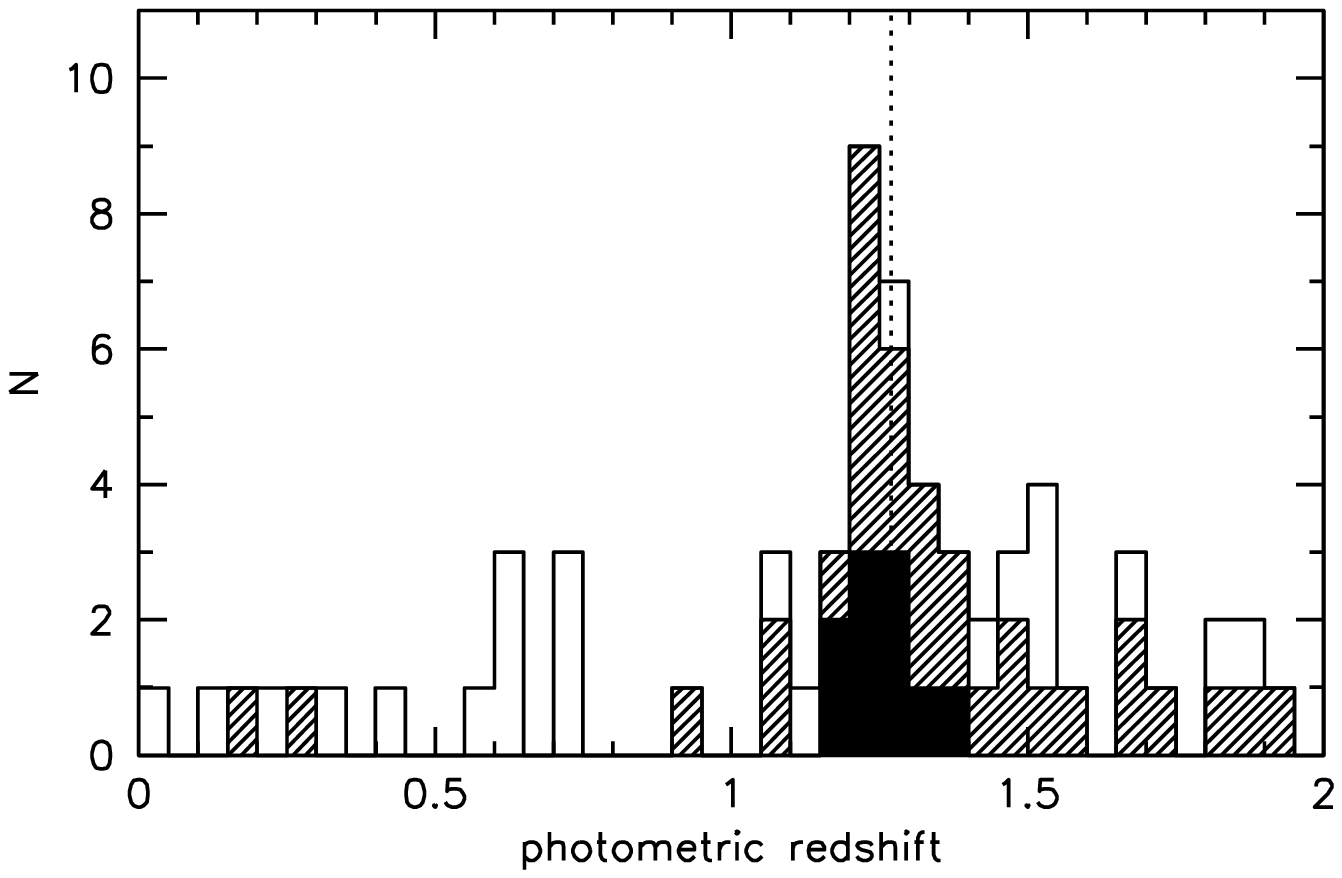}}
\figcaption{\small
Photometric redshift distribution in the HST field. The dotted line
indicates the spectroscopic redshift of
the cluster, $z=1.27$. This redshift
coincides with a strong peak in the photometric redshift distribution.
The solid black histogram shows the photometric redshift
distribution of spectroscopically
confirmed members, and
the hatched histogram shows galaxies without spectroscopic redshift.
The open histogram shows confirmed field galaxies.
\label{zhis.plot}}
\end{center}}

Based on the photometric redshift distribution of confirmed members
we assume that galaxies with
$1.14<z_{\rm phot} <1.40$ are cluster
members. One of these is a luminous AGN at $z_{\rm spec}=0.889$
which we discard. The final sample of 25 likely cluster
galaxies includes the ten confirmed
members.

\begin{figure*}
\epsfxsize=16cm
\epsffile{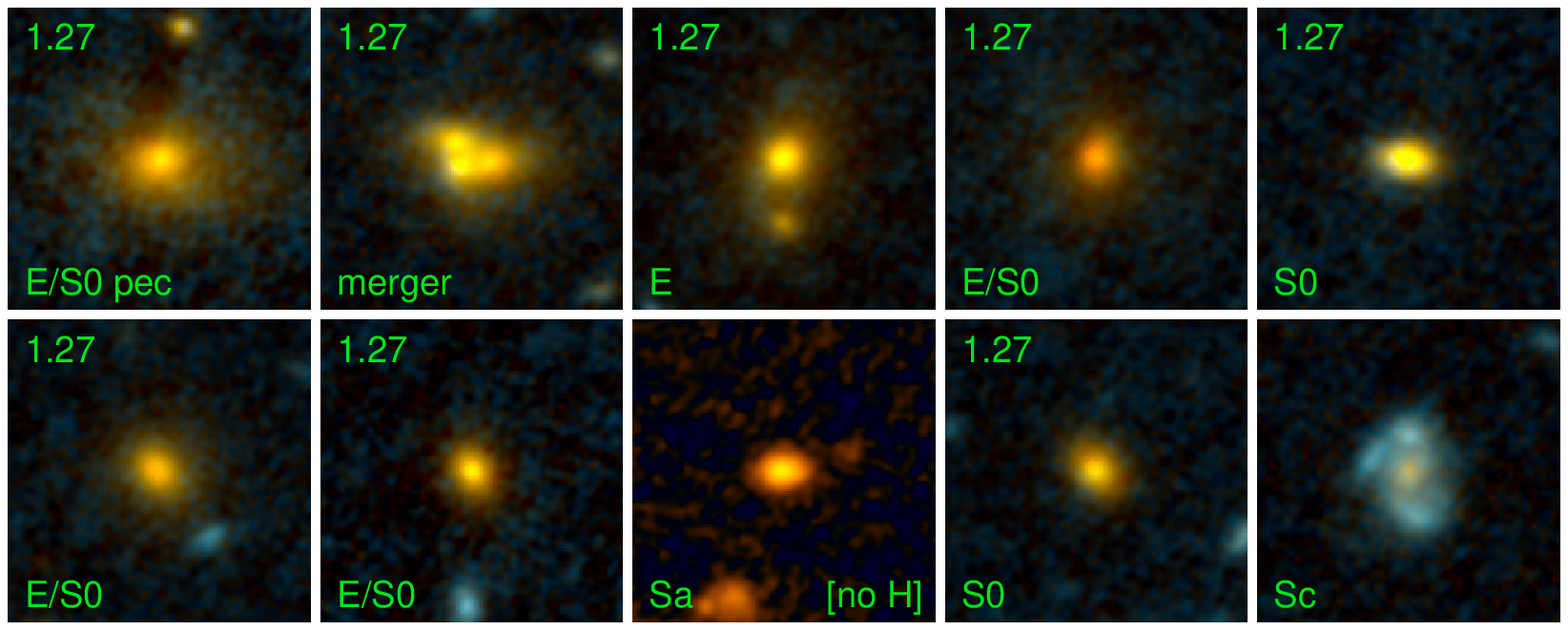}
\caption{\small
Color images of the ten brightest (in rest frame $V$) galaxies in \j0848,
created from the NIC3 $H_{F160W}$ and WFPC2 $I_{F814W}$ exposures.
The WFPC2 images were smoothed to the NIC3 resolution of
$0\farcs 22$ FWHM.
Each image measures
$5\farcs 1 \times 5\farcs 1$ ($53 \times 53$\,\h50min\,kpc).
The galaxy labeled ``no H'' was only observed in $I_{F814W}$.
Spectroscopically confirmed cluster members are labeled ``1.27''.
\label{stamps.plot}}
\end{figure*}

\section{Morphologies}

Morphological classifications of
all galaxies in the WFPC2 and/or NIC3 field to $K_s=20.6$ were performed
by P.\ G.\ v.\ D., S.\ A.\ S., and B.\ P.\ H.,
following the method outlined in Fabricant, Franx, \& van Dokkum (2000).
We preferentially used the deeper $H_{F160W}$ images for galaxies that
were observed with both WFPC2 and NIC3. 
The reliability of such
visual classifications has been discussed extensively
(e.g., Smail et al.\ 1998, Fabricant et al.\ 2000).
In general, when comparing independent sets of classifications
the early-type galaxy fraction is robust, but
there is large scatter in the relative numbers of elliptical
and S0 galaxies
(e.g., Fabricant et al.\ 2000, van Dokkum et al.\ 2000).
In the present study
we do not discuss the fractions of elliptical galaxies
and S0 galaxies separately.

Three of the 25 likely cluster members could not be classified;
they were discarded from the sample to leave 22 probable
cluster members with morphological classifications.
Eleven of the 22 (50\,\%) are E or S0 galaxies, nine (41\,\%) are
spirals, and two (9\,\%) are classified as ``merger/peculiar''.
Both ``merger/peculiar'' galaxies are spectroscopically confirmed
cluster members. In nearby clusters, the morphological fractions are a
strong function of the local galaxy density (Dressler 1980).
Following the prescription of Dressler (1980) we find that the
logarithm of the galaxy density $\log \rho_{\rm proj} \sim
1.3\,h_{50}^{-2}$\,Mpc$^{-2}$ for \j0848.  The predicted early-type
fraction from the local morphology-density relation is $70$\,\%,
higher than the observed fraction of $50^{+17}_{-12}$\,\%,
although the difference is not significant. Color images of the
ten brightest galaxies in \j0848{} are shown in Fig.\
\ref{stamps.plot}, along with their morphological classification.

\section{A Luminous Triple Merger at $z=1.27$}

The second brightest cluster galaxy is a single object in ground based
images, but is resolved into three individual galaxies in the HST
images (cf.\ Fig.\ \ref{stamps.plot}). The three objects have very
similar luminosities, morphologies, and colors, and are separated from
each other by only $0\farcs 6$ ($\approx
6$\,\h50min\,kpc). The probability of finding
three unrelated objects with such a small separation and high luminosity in
our sample is $\sim 10^{-7}$.  The probability that two are physically
associated and one is unrelated is $\sim 10^{-4}$.  Hence
we conclude that the three objects are bound, and are
in the late stages of a triple merger.  This interpretation is
confirmed by the presence of tidal tails.
The three galaxies will probably evolve into a $\sim 3L_*$ elliptical
galaxy in $< 500$\,Myr (e.g., Rix \& White 1989).
Note that we have counted the three merging galaxies as one object
in determining the fraction of ``merger/peculiar'' galaxies
in \j0848.

The integrated spectrum of the three objects shows no evidence for
significant star formation (Stanford et al.\ 1997).
Such apparently dissipationless mergers were previously
observed in the outskirts of the cluster MS\,1054--03 at $z=0.83$
(van Dokkum et al.\ 1999).
Similar to MS\,1054--03, \j0848{} is not
relaxed (Stanford et al.\ 2000). The triple merger may occur in a subclump,
where the relative velocities of
galaxies are sufficiently low. If the merging galaxies were
accreted from the field
a major question is how and when they
lost their gas.

Interestingly, the brightest and third brightest
cluster galaxy also show evidence for interactions. The brightest
cluster galaxy has an extended, asymmetric envelope. Such
features can be explained by a recent merger or accretion event
(e.g., Statler, Smecker-Hane, \& Cecil 1996). The third brightest
galaxy has a close companion, at a distance of $1\farcs 1$.
The probability of finding two unrelated galaxies within
this distance in our sample is $\sim 8$\,\%, and
the colors of the companion object suggest that it is a
cluster member. Spectroscopy of the companion galaxy is
needed to confirm that the galaxies are physically associated.

\section{Color-Magnitude Relation}

The color-magnitude (CM) relation of probable members of \j0848{} is
shown in Fig.~\ref{cm.plot}. Rest frame $U-V$ colors and
absolute $V$ magnitudes were calculated from the observed (ground
based) $I$ and
$J$ band magnitudes (see, e.g., van Dokkum
et al.\ 2000). Colors were measured in $3''$ apertures.
The broken line indicates the CM relation of
early-type galaxies in Coma (Bower, Lucey, \& Ellis 1992).

Early-type galaxies in \j0848{} follow a narrow CM relation, with the
exception of a few faint outliers.  The CM relation is noticably
flatter than in the Coma cluster; a fit to the early-type galaxies
with $M_V^T<-22.5$ gives a slope of $-0.02 \pm 0.03$, compared to
$-0.08 \pm 0.01$ for Coma (Bower et al.\ 1992). This flat slope
is probably not caused by the effects of
color gradients, since these are expected to steepen the slope of
the CM relation (Scodeggio 2001).

As a result of the difference in slope, the offset of the CM relation
from that of the Coma cluster depends on magnitude. At $M_V^T =
-23$ the offset is $0.3 \pm 0.1$ magnitudes, consistent with
predictions of stellar population synthesis models for populations
formed at $z \gtrsim 2$ (e.g., Worthey 1994). The constraints are
not very strong because of the relatively large uncertainty in
the offset. The corresponding rest
frame $V$ band luminosity evolution is $\sim 1$ magnitude,
consistent with the evolution of the $M/L_B$ ratio and
the luminosity function to $z \sim 1$ (van Dokkum et al.\ 1998; De
Propris et al.\ 1999).

The scatter was determined using the biweight statistic (Beers, Flynn,
\& Gebhardt 1990). The observed scatter among the early-type galaxies
is only $0.06^{+0.04}_{-0.03}$ in (ground based) rest frame $U-V$
color, identical to the expected scatter from measurement errors
alone. Ten of the early-type galaxies are observed with NIC3 and
WFPC2, allowing us to obtain an independent measurement of the
scatter.  The observed scatter is $0.10^{+0.07}_{-0.05}$ in $I_{F814W}
- H_{F160W}$.  After correcting for a
measurement error of $0.07$ magnitudes this gives an intrinsic scatter
of $\approx 0.07$ in rest frame $U-R$, or $\approx 0.04$ in
$U-V$. This scatter is very similar to that in the nearby
Coma cluster ($0.047 \pm 0.005$ in $U-V$; De Propris et al.\ 2000).
This result extends to $z\sim 1.3$ the results of
Stanford et al.\ (1998) on the CM relation at $z<1$.

\vbox{
\begin{center}
\leavevmode
\hbox{%
\epsfxsize=7cm
\epsffile{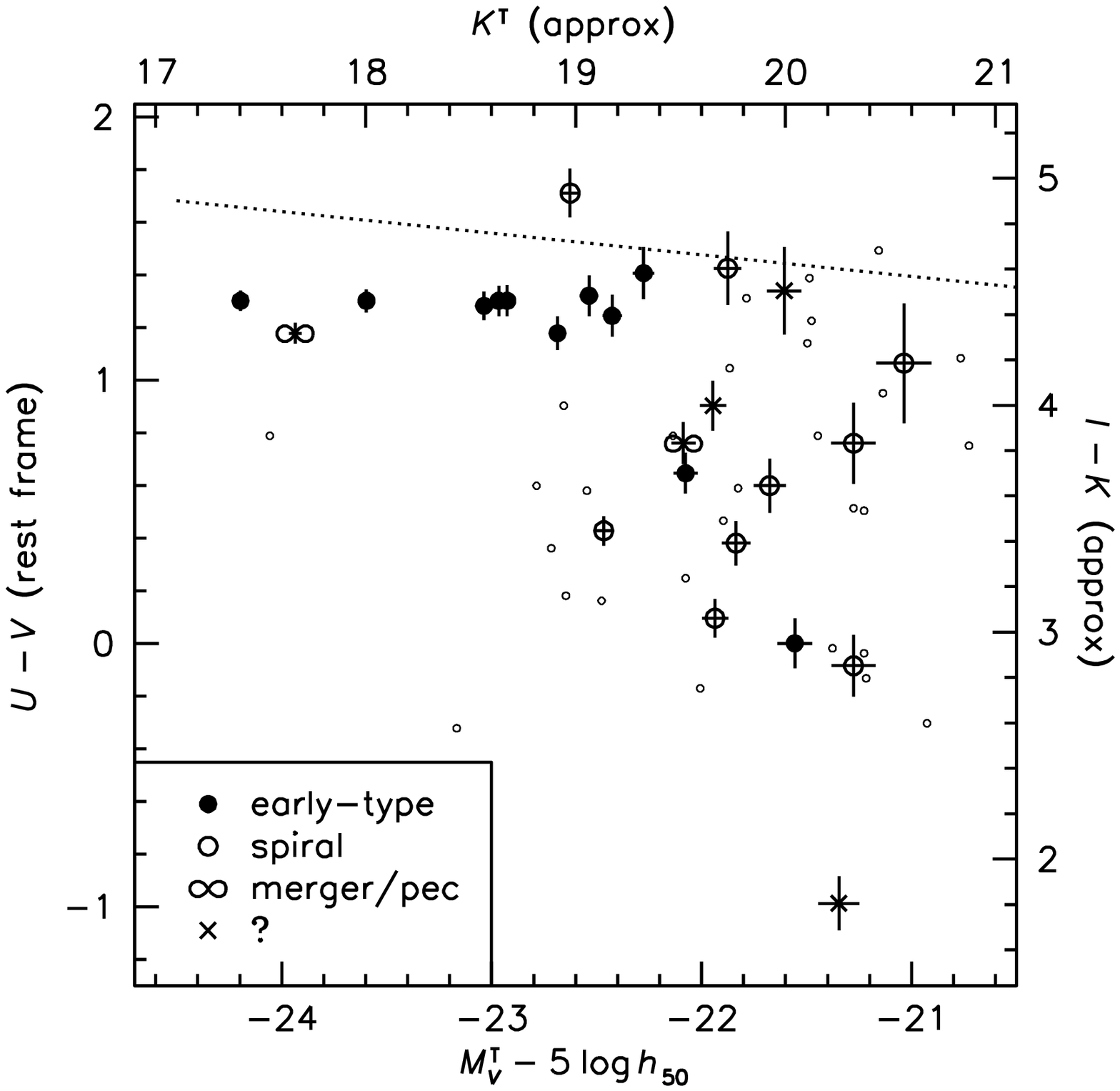}}
\figcaption{\small
Rest frame $U-V$ versus $V$ color-magnitude relation of \j0848{} for
the sample of 25 probable cluster members.
Small open symbols are non-members
as determined from their photometric redshifts. Spectroscopically
confirmed field galaxies are not shown.
One sigma errorbars are shown for probable cluster members
only. The dotted line
is the CM relation of Coma, as determined by Bower, Lucey, \& Ellis
(1992). The CM relation of \j0848{} appears to be
less steep than that of Coma.
\label{cm.plot}}
\end{center}}

Spiral galaxies show a large scatter in their colors: the majority are
much bluer than the CM sequence defined by the early-type galaxies,
but a few are very red. The triple merger
is only $0.12 \pm 0.02$ magnitudes bluer in rest frame $U-V$ than
early-type galaxies of the same luminosity.  This system resembles the
red mergers seen in the cluster MS\,1054--03 at $z=0.83$ (van Dokkum
et al.\ 1999).

\section{Discussion}

%The montage in Fig.\ \ref{stamps.plot} shows directly that a 
%significant fraction of the most luminous galaxies
%in clusters was still in the process of formation at $z=1.27$:
%only one of the five most luminous galaxies in \j0848{} appears to
%be a normal, unperturbed early-type galaxy.

The presence of red mergers in high redshift clusters is direct evidence
against formation of all massive elliptical galaxies in a ``monolithic
collapse'' at very high redshift, and in qualitative agreement with
hierarchical galaxy formation models. The triple merger in \j0848{}
shares many of its characteristics with the luminous red mergers seen in
MS\,1054--03 at $z=0.83$ (van Dokkum et al.\ 1999). Despite its higher
redshift, the merger fraction in \j0848{} may not be as high as in
MS\,1054--03.  The mergers in MS\,1054--03 occur preferentially in
the outskirts of the cluster, and it would be interesting to extend
the study of \j0848{} to larger radii. The merger fraction may be
correlated with the dynamical state of clusters; this may be tested by
studying a sample of clusters in different stages of collapse.

\vbox{
\begin{center}
\leavevmode
\hbox{%
\epsfxsize=7cm
\epsffile{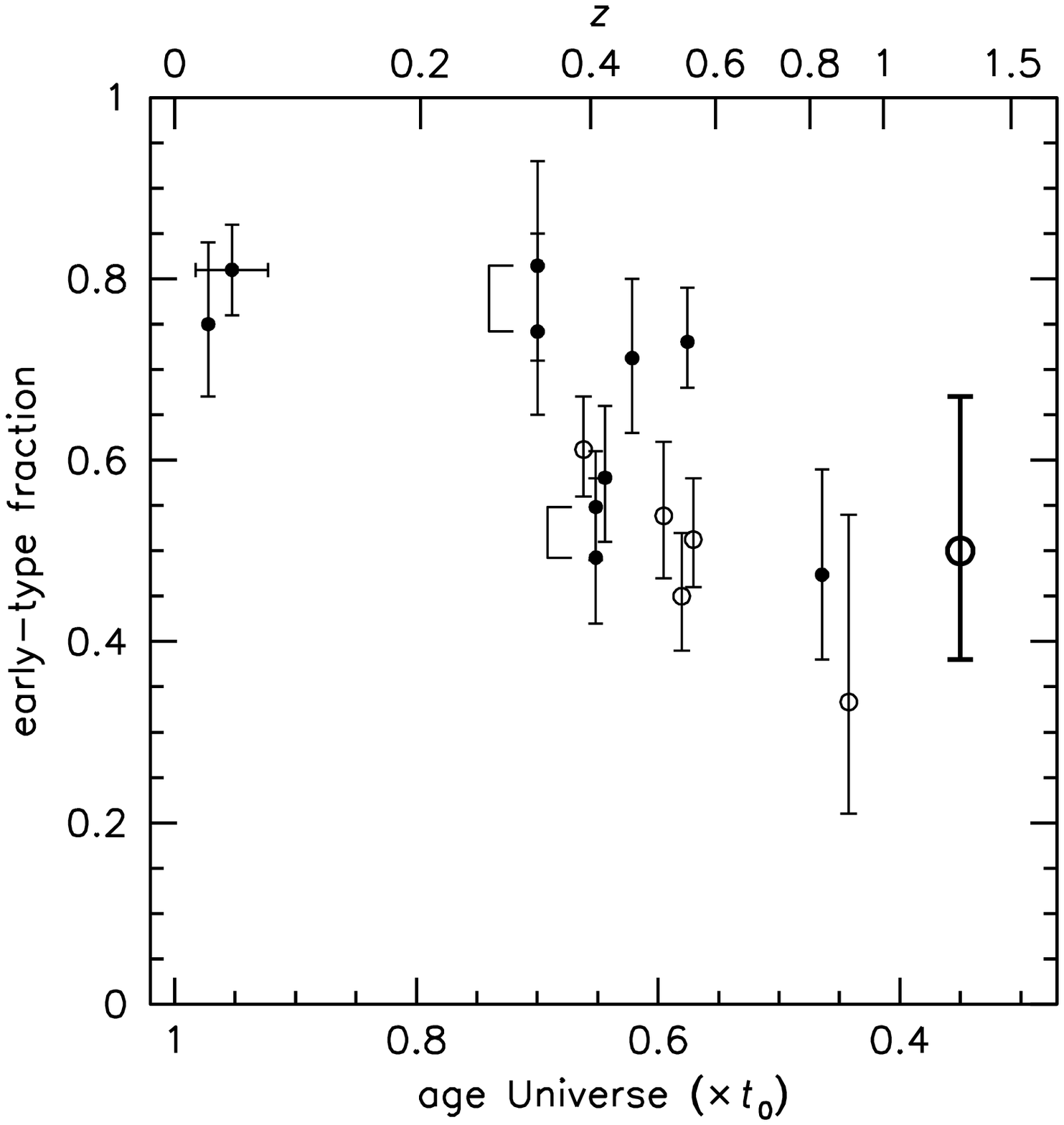}}
\figcaption{\small
Evolution of the early-type galaxy fraction in clusters. Data points
at $z<1$ are taken from the literature (see text).
Solid symbols are clusters with $L_X>10^{44.5}$
ergs\,s$^{-1}$, and open symbols indicate $L_X<10^{44.5}$
ergs\,s$^{-1}$.  The large symbol
at $z=1.27$ is \j0848 which has $L_X = 0.7 \times 
10^{44}$ ergs\,s$^{-1}$ (Stanford
et al.\ 2000). The early-type galaxy fraction in \j0848{} is similar
to that of clusters at $0.5<z<1$.
\label{evoearly.plot}}
\end{center}}

The fraction of early-type galaxies provides a complementary
measurement of morphological evolution. 
Figure \ref{evoearly.plot} shows the evolution of the early-type
galaxy fraction from $z=0$ to $z=1.3$. Data points at $z<1$ are taken
from Dressler et al.\ (1997), Andreon et al.\ (1997), Fabricant et
al.\ (2000), Lubin et al.\ (1998), and van Dokkum et al.\ (2000).
The early-type fraction in \j0848{} is similar to that
in clusters at $0.5<z<1$.  The interpretation is complicated by the
effects of infall of field galaxies
onto clusters between $z=1.27$ and $z=0$ (e.g.,
Ellingson et al.\ 2001).
Furthermore, clusters at $z \approx 1.3$
may be rare and not representative of the progenitors of nearby
clusters such as Coma (e.g., Kauffmann 1995). As a result of these
effects the observed evolution of the early-type galaxy fraction
may underestimate the true evolution.

The low and constant scatter in the CM relation from $z=0$ to $z=1.3$
can be interpreted as evidence that early-type galaxies formed at very
high redshift (e.g., Ellis et al.\ 1997). However, it is also
consistent with models in which early-type galaxies are continuously
transformed from spiral galaxies (van Dokkum \& Franx 2001).  In
that case, the low scatter implies that the
morphological transformations are probably
not accompanied by strong star
bursts.  The red colors of the triple merger in \j0848, and of spiral
galaxies and mergers in clusters at lower redshift (e.g., Poggianti et
al.\ 1999, van Dokkum et al.\ 2000), are consistent with this
scenario.

The flattening of the slope of the CM relation with redshift, if
confirmed for other clusters, may indicate that the most massive
early-type galaxies harbor young stellar populations at $z\sim 1.3$,
consistent with predictions of semi-analytical models for galaxy
formation (e.g., Kauffmann \& Charlot 1998). However, a
similar effect is expected in single age
formation models with a galactic wind (Arimoto
\& Yoshii 1987).  In these models the colors of massive
galaxies evolve more rapidly because they have higher metallicities.
Measurements of line indexes and $M/L$ ratios of
the most massive early-type galaxies in \j0848{} may help to determine
the ages of the stars in these galaxies.

\acknowledgements{
We thank Chris Hanley for help with the WFPC2 and NICMOS image
reductions.
P. G. v. D. acknowledges support by NASA through Hubble Fellowship
grant HF-01126.01-99A
awarded by the Space Telescope Science Institute, which is
operated by the Association of Universities for Research in
Astronomy, Inc., for NASA under contract NAS 5-26555.
Support for S. A. S. came from NASA/LTSA grant NAG5-8430,
and STScI grants GO-06812.02 and GO-07872.01. B. H. acknowledges
support from NASA/Chandra grant GO0-1082A.
Portions of this research were carried out
by the Jet Propulsion Laboratory, California 
Institute of Technology, under a contract with NASA.
}

\end{document}